\documentclass[showpacs,twocolumn,aps,pra,superscriptaddress]{revtex4}
\usepackage{txfonts}
\usepackage{amsfonts}
\usepackage{epsfig}
\usepackage{subfigure}

\usepackage{graphicx}
\usepackage{dcolumn}
\usepackage{bm}
\begin{document}

\title{Experimental study on discretely modulated continuous-variable quantum key distribution}

\author{Yong Shen}

\author{Hongxin Zou}\email{hxzou@nudt.edu.cn}
\affiliation{Department of Physics, The National University of
Defense Technology, Changsha 410073, PR China}
\author{Liang Tian}
\affiliation{Department of Physics, The National University of
Defense Technology, Changsha 410073, PR China} \affiliation{College
of Optoelectronic Science and Engineering, The National University
of Defense Technology, Changsha 410073, PR China}
\author{Pingxing Chen}
\affiliation{Department of Physics, The National University of
Defense Technology, Changsha 410073, PR China}
\author{Jianmin Yuan}
\affiliation{Department of Physics, The National University of
Defense Technology, Changsha 410073, PR China}

\begin{abstract}
We present a discretely modulated continuous-variable quantum key
distribution system in free space by using strong coherent states.
The amplitude noise in the laser source is suppressed to the
shot-noise limit by using a mode cleaner combined with a frequency
shift technique. Also, it is proven that the phase noise in the
source has no impact on the final secret key rate. In order to
increase the encoding rate, we use broadband homodyne detectors and
the no-switching protocol. In a realistic model, we establish a
secret key rate of 46.8 kbits/s against collective attacks at an
encoding rate of 10 MHz for a 90\% channel loss when the modulation
variance is optimal.

\end{abstract}

\maketitle
\section{INTRODUCTION}
Continuous-variable quantum key distribution (CV-QKD) by using
coherent states \cite{1} was introduced as an alternative to the
single-photon-based discrete quantum key distribution (QKD) protocol
\cite{2}. In this protocol, two legitimate users (Alice and Bob) use
coherent states whose X and P quadratures are Gaussian modulated to
establish a shared secret key. CV-QKD has made great achievements
during the past few years. At first, it was thought that no secret
key rate could be obtained when the channel loss was larger than 3
dB, subsequently, the 3-dB loss limit was beaten by the methods of
reverse reconciliation \cite{3} and was experimentally demonstrated
\cite{4}. At the same time, another method called postselection
\cite{5} was proposed, which can also beat the 3-dB loss limit. Just
like the discrete QKD protocol, at first, it was believed that the
security of CV-QKD was based on the random switching of bases that
Bob measures. Subsequently, it was found that, without switching,
CV-QKD is also secure \cite{6}, and has performed experimental
demonstrations \cite{7,8}. Several experiments of Gaussian-modulated
CV-QKD with optical fibers have been implemented \cite{9,10,11}.
However, the distance between Alice and Bob is much shorter than
that in the discrete QKD because the reconciliation efficiency of
continuous variables is much lower than that of discrete variables
when the signal-to-noise ratio (SNR) is small.

In order to adapt CV-QKD for long-distance communication, two CV-QKD
protocols with discrete modulation were proposed \cite{21,12}, and
the former was recently experimentally implemented \cite{13}. In
these schemes, instead of Gaussian modulation, Alice modulates the
quadratures of coherent states discretely. In the former protocol,
Eve's attacks were restricted by tomography performed by Bob,
whereas the latter offered unconditional proofs with no assumption.
The security of the latter scheme is guaranteed by the optimality of
Gaussian attacks \cite{14,15}, that is, when the covariance matrix
of the state shared by Alice and Bob is the same as in the Gaussian
modulated case, the secret key rate is also the same. However,
discretely modulated CV-QKD is secure only when the modulation
variance is small. So the SNR in discretely modulated CV-QKD is much
smaller than that of the Gaussian modulated CV-QKD experiments so
far, which means that the noise is more fatal in discretely
modulated CV-QKD. The noise in CV-QKD mainly consists of two parts,
quantum noise and classical noise. The former is induced by channel
loss and can not be suppressed; the latter is called excess noise
and can be arbitrarily suppressed in principle.

In this paper, we present an experimental implementation of
discretely modulated CV-QKD described in Ref.\cite{12} in free
space. In order to increase the bandwidth of cryptography and to
remove the noise caused by imprecise control of the relative phase
between the signal and the local oscillator (LO), we use the
no-switching protocol \cite{6} instead of randomly switching the
quadrature that Bob measures. Unlike Ref. \cite{13}, we use strong
coherent states instead of weak coherent states for obtaining a
sizable feedback signal to lock the relative phase. Additionally,
the amplitude noise in the laser source is suppressed at the
shot-noise limit by using a mode cleaner combined with a frequency
shift technique. Also, it is proven that the phase noise in the
source has no impact on the final secret key rate. As a result, we
establish a secret key rate of 46.8 kbits/s for a 90\% lossy channel
(which corresponds to a 50 km standard telecom fiber with a
0.2-dB/km loss) under the realistic model \cite{4}.

\section{THEORETICAL EVALUATION OF THE SECRET KEY RATES}
\subsection{PROTOCOL DESCRIPTION}
In this section, we detail the calculation of the security bound of
discretely modulated CV-QKD under collective attacks by considering
the noise in the laser source.

When the noise in the source is not considered, the protocol runs as
follows \cite{12}. (i) Alice prepares one of the four coherent
states: $ \left| {\alpha _k } \right\rangle  = \left| {\alpha
e^{i(2k + 1)\pi /4} } \right\rangle $ with $k \in \{ 0,1,2,3\}$ and
sends it to Bob. The real number $\alpha$ is chosen so as to
maximize the secret key rate. (ii)When Bob receives the state, he
uses a 50:50 beamsplitter to split the state into two beams. Then he
measures the $X$ quadrature of one beam and the $P$ quadrature of
the other. (iii) The signal of the modulated and measured value
encodes the bit of the raw key, so after Bob's measurement, he and
Alice share correlated strings of bits. By reconciliation and
privacy amplification they can achieve secret key. This is the $\it
{prepare\ and\ measure}$ version of the protocol. It is easy to
implement experimentally in this version, but difficult to analyze
theoretically. Usually the security is established by considering
the equivalent entanglement-based scheme. In this scheme, Alice has
a pure two-mode entanglement state \cite{12}
\begin{equation}\label{1}
\left| {\Phi _{AB_0 } } \right\rangle  = \frac{1}{2}\sum\limits_{k =
0}^3 {\left| {\psi _k } \right\rangle _A \left| {\alpha _k }
\right\rangle _{B_0 } } , \end{equation}
 where the states
\begin{equation}\label{2}
\left| {\psi _k } \right\rangle  = \sum\limits_{m = 0}^3
{\frac{1}{2}e^{ - i(1 + 2k)m\pi /4} \left| {\phi _m } \right\rangle
}
\end{equation}
are orthogonal with each other. The state $\left| {\phi _m }
\right\rangle$ is defined as follows:
\begin{equation}\label{3}
\left| {\phi _m } \right\rangle  = \frac{{e^{ - \alpha ^2 /2}
}}{{\sqrt {\xi _m } }}\sum\limits_{n = 0}^\infty  {\frac{{\alpha
^{4n + m} }}{{\sqrt {(4n + m)!} }}} ( - 1)^n \left| {4n + m}
\right\rangle,
\end{equation}
where
\begin{equation}\label{4}
\begin{array}{l}
 \xi _{0,2}  = {\textstyle{1 \over 2}}\exp \left( { - \alpha ^2 } \right)\left( {\cosh \left( { - \alpha ^2 } \right) \pm \cos \left( { - \alpha ^2 } \right)} \right), \\
 \xi _{1,3}  = {\textstyle{1 \over 2}}\exp \left( { - \alpha ^2 } \right)\left( {\sinh \left( { - \alpha ^2 } \right) \pm \sin \left( { - \alpha ^2 } \right)} \right). \\
 \end{array}
\end{equation}
Alice holds mode $A$ and sends mode $B_0$ to Bob. Then she uses a
set of projection operators $\left| {\psi _k } \right\rangle
\left\langle {\psi _k } \right| $ ($k=0,1,2,3$) to measure the mode
she keeps. If mode $A$ collapses into $\left| {\psi _k }
\right\rangle$, then the mode sent to Bob collapses into $\left|
{\alpha _k } \right\rangle$.

\subsection{THE ENTANGLEMENT-BASED SCHEME FOR A NOISY SOURCE}
When the noise in the source is taken into account, things are
slightly different. In the $\it{prepare\ and\ measure}$ scheme, due
to the noise in the laser source and the imperfection of modulation,
the state Alice sends to Bob is not a pure state $ \left| {\alpha _k
} \right\rangle$, instead it is a noisy mixed state $\rho _{B_0 }^k
$ Without loss of generality, let us assume that the noise on the
$X$ quadrature and the $P$ quadrature have the same variance $\delta
\varepsilon$, and their mean values are both zero. Also, we assume
that the noise is induced by a neutral person Fred, and the
eavesdropper Eve can not benefit from it.

\begin{figure}
\centering
\includegraphics[width=.5\textwidth]{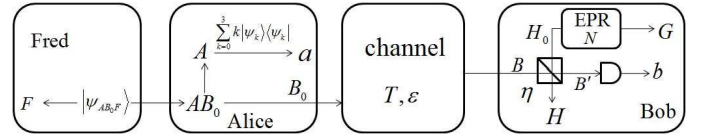}
\caption{The entanglement-based scheme of discretely modulated
CV-QKD when considering the noise in the source. Bob uses heterodyne
detection and it is assumed that Eve can not benefit from the
imperfection of Bob's detector}\label{pic1}
\end{figure}

In the equivalent entanglement-based scheme, as shown in
Fig.\ref{pic1}, Fred has a pure three-mode entanglement state,
\begin{equation}\label{5}
 \left| {\Phi _{AB_0 F} } \right\rangle  =
\frac{1}{4}\sum\limits_{k = 0}^3 {\left| {\psi _k } \right\rangle_A
\left| {\varphi _{B_0 F}^k } \right\rangle } ,
\end{equation}
where $\left| {\varphi _{B_0 F}^k } \right\rangle $ satisfies $ tr_F
\left( {\left| {\varphi _{B_0 F}^k } \right\rangle \left\langle
{\varphi _{B_0 F}^k } \right|} \right) = \rho _{B_0 }^k $. Fred
keeps mode $F$ and sends modes $A$ and $B_0$ to Alice. Alice holds
mode $A$ and sends mode $B_0$ to Bob. Then, she uses a set of
projection operators $\left| {\psi _k } \right\rangle \left\langle
{\psi _k } \right| $ ($k=0,1,2,3$) to measure the mode she keeps. If
mode $A$ collapses into $\left| {\psi _k } \right\rangle$, the mode
sent to Bob collapses into $\rho _{B_0 }^k $.
On Bob's side, we consider the realistic model \cite{4}, in which
Eve cannot benefit from the noise added by Bob's detector. In the
entanglement-based scheme we can simplify the description of the
realistic detector on Bob side \cite{10}. As shown in
Fig.\ref{pic1}, the inefficiency of Bob's detector is modeled by a
beam splitter with transmission $\eta$, while the electronic noise
$\upsilon$ of Bob's detector is modeled by a thermal state
$\rho_{H_0}$ with variance $N$, which enters the other input port of
the beam splitter. Then, Bob uses a perfect heterodyne detector to
measure state $\rho_{B'}$. It is obvious to obtain that the variance
of the result of Bob's measurement $V_b$ is \cite{25}
\begin{equation}\label{38}
V_b  = \frac{\eta }{2}V_B  + \frac{{1 - \eta }}{2}N + \frac{1}{2},
\end{equation}
where $V_B$ is the variance of state $\rho_B$. Since the detector's
efficiency is $\eta$ and the electronic noise is $\upsilon$, we can
also obtain that
\begin{equation}\label{39}
V_b  = \eta \left( {\frac{{V_B }}{2} + \frac{1}{2}} \right) + \left(
{1 - \eta } \right) + \upsilon ,
\end{equation}
so we can obtain $N = 1 + {{2\upsilon } \mathord{\left/
 {\vphantom {{2\upsilon } {(1 - \eta )}}} \right.
 \kern-\nulldelimiterspace} {(1 - \eta )}}$. To consider the thermal state $\rho_{H_0}$ as the reduced state
obtained from a two-mode Gaussian state $\rho_{GH_0}$ of variance
$N$ allows us to simplify the calculations.

\subsection{THE OPTIMALITY OF GAUSSIAN ATTACKS FOR A NOISY SOURCE}
In collective attacks, Eve uses an ancilla to interact with each
pulse that Alice sends to Bob. After the interaction, the global
state $\rho_{AB_0F}$ turns into $\rho_{ABEF}$. On Bob's side, before
his measurement, the received pulse interferes with the thermal
state $\rho_{H_0}$, and the global state becomes $\rho_{AB'EFGH}$
Under collective attacks, when Alice and Bob use reverse
reconciliation and the reconciliation efficiency is $\beta$, the
secret key rate is \cite{12}
\begin{equation}\label{6}
K = \beta I(a:b) - \chi (b:E),
\end{equation}
where $a$, $b$ represent the classical data of Alice and Bob,
$I(a:b)$ is the Shannon mutual information between $a$ and $b$, and
$\chi(b:E)$ is the Holevo bound, an upper bound for Eve's accessible
information. When considering both the noisy source and the
realistic detector, it is rather complicated to derive the
information that Eve gets by using the method introduced in Ref.
\cite{23}, since the global state is an eight-mode state. So, we do
not derive the secret key rate directly. Instead, we find a lower
bound to $K$ \cite{18}
\begin{equation}\label{7}
\tilde K = \beta I(a:b) - \chi \left( {b:EF} \right),
\end{equation}
and $\tilde K \le K$ always holds. It is obvious that when the noise
in the source is small, $\tilde K$ will be very close to $K$.
Additionally, when the source is noiseless, $\tilde K =K$ holds. The
Holevo bound $\chi\left(b:EF\right)$ is defined as
\begin{equation}\label{21}
\chi \left( {b:EF} \right) = S\left( {\rho _{EF} } \right) - \int
{p(b)S\left( {\rho _{EF}^b } \right)db} ,
\end{equation}
where $p(b)$ is the probability of the result of Bob's measurement.
Before Bob's measurement, the global state is $\rho_{ABEF}$, so we
obtain $S\left( {\rho _{EF} } \right) = S(\rho _{AB} ) $. After
Bob's measurement, the global state comes into $\rho_{AEFGH}^b$,
thus we have ${S\left( {\rho _{EF}^b } \right) = S\left( {\rho
_{AGH}^b } \right)}$.

Notice that state $\rho_{AB'GH}$ is determined by state $\rho_{AB}$,
so $K$ is a function of $\rho_{AB}$. According to the optimality of
Gaussian attacks \cite{14,15}, for all the two-mode states
$\rho_{AB}$ with the same covariance matrix, $\tilde K(\rho_{AB})$
achieves the minimum value when $\rho_{AB}$ is Gaussian. In the
following, instead of $K$, we will derive its lower bound $\tilde
K$. When the channel's transmittance is $T_0$ and the excess noise
is ${\varepsilon _0 }$, the variance matrix of $\rho_{AB}$ is
\begin{equation}\label{8}
\gamma _{AB}  = \left[ {\begin{array}{*{20}c}
   {\left( {V_A  + 1} \right)I_2 } & {\sqrt {T_0 } Z\sigma _Z }  \\
   {\sqrt {T_0 } Z\sigma _Z } & {\left[ {T_0 \left( {V_A  + \varepsilon _0  + \delta \varepsilon } \right)} \right]I_2 }  \\
\end{array}} \right],
\end{equation}
where $Z = 2\alpha ^2 \left( {\xi _0^{{\textstyle{3 \over 2}}} \xi
_1^{ - {\textstyle{1 \over 2}}} + \xi _1^{{\textstyle{3 \over 2}}}
\xi _2^{ - {\textstyle{1 \over 2}}}  + \xi _2^{{\textstyle{3 \over
2}}} \xi _3^{ - {\textstyle{1 \over 2}}} + \xi _3^{{\textstyle{3
\over 2}}} \xi _0^{ - {\textstyle{1 \over 2}}} } \right) $ reflects
the correlation between mode $A$ and mode $B$, $V_A = 2\alpha ^2$ is
just the modulation variance in the $\it{prepare\ and\ measure}$
scheme, $I_2$ is a two-dimensional unit matrix and $\sigma _z =
diag\left( {1, - 1} \right) $. Then, we will represent the
corresponding case in the Gaussian modulated protocol. In this case,
Alice modulates the pure coherent states with Gaussian variables,
whose variance is $V_A$. Then she sends them to Bob via a channel
with transmittance $T$ and excess noise $\varepsilonup$. In the
equivalent entanglement-based scheme, the variance of $\rho_{AB}$ is
\begin{equation}\label{9}
\gamma _{AB}^G  = \left[ {\begin{array}{*{20}c}
   {\left( {V_A  + 1} \right)I_2 } & {\sqrt T Z_{EPR} \sigma _z }  \\
   {\sqrt T Z_{EPR} \sigma _z } & {\left[ {T\left( {V_A  + \varepsilon } \right) + 1} \right]I_2 }  \\
\end{array}} \right],
\end{equation}
where $Z_{EPR}  = \sqrt {V_A^2  + 2V_A }$ \cite{12}. The
entanglement states used in Gaussian-modulated CV-QKD are maximally
correlated, while those in discretely modulated CV-QKD are not, so
$Z < Z_{EPR} $. To make $\gamma _{AB}^G$  equal to $ \gamma _{AB} $,
we get
\begin{equation}\label{20}
T = T_0 \frac{{Z^2 }}{{Z_{EPR}^2 }},\quad\varepsilon  =
\frac{{Z_{EPR}^2 }}{{Z^2 }}\left( {V_A  + \varepsilon _0  + \delta
\varepsilon } \right) - V_A .
\end{equation}
According to the optimality of Gaussian attacks \cite{14,15}, if we
use discrete modulation, when the modulation variance is $V_A$, the
variance of the noise in the source is $\delta\varepsilon$, the
channel's transmittance is $T_0$ and excess noise is
$\varepsilon_0$, the lower bound of the secret key rate is just the
same as the secret key rate of the case in which Alice uses Gaussian
modulation with variance $V_A$, and the channel's transmittance and
excess noise are given by Eq. (\ref{20}).

\subsection{CALCULATION OF THE SECRET KEY RATE}
We call $\chi _c  = {1 \mathord{\left/
 {\vphantom {1 T}} \right.
 \kern-\nulldelimiterspace} T} - 1 + \varepsilon$ the noise added by the channel, and $
\chi _d  = 2{{\left( {1 + \upsilon } \right)} \mathord{\left/
 {\vphantom {{\left( {1 + \upsilon } \right)} \eta }} \right.
 \kern-\nulldelimiterspace} \eta } - 1
$ the noise induced by the heterodyne detector. Then the total noise
added between Alice and Bob is
\begin{equation}\label{18}
\chi _t  = \chi _c  + {{\chi _d } \mathord{\left/
 {\vphantom {{\chi _d } T}} \right.
 \kern-\nulldelimiterspace} T}.
\end{equation}
When Bob uses heterodyne detection, the mutual information between
Alice and Bob is \cite{19,24}
\begin{equation}\label{10}
I\left( {a:b} \right) = \log _2 \left( {\frac{{V + \chi_t}}{{\chi_t
+ 1}}} \right),
\end{equation}
where $V=V_A+1$.

Then we will derive $S\left(\rho_{AB}\right)$. Let $\alpha = V $,
$\beta = T(V + \chi _c ) $, and $\gamma = \sqrt {T(V^2 - 1)} $, the
symplectic eigenvalues $\lambda_1$, $\lambda_2$ satisfy
\begin{equation}\label{22}
\begin{array}{l}
 \lambda _1^2  + \lambda _2^2  = \alpha^2 + \beta^2 - 2\gamma^2, \\
 \lambda _1^2 \lambda _2^2  = \left(\alpha \beta - \gamma^2\right)^2 . \\
 \end{array}
\end{equation}
So we obtain
\begin{equation}\label{23}
\lambda _{1,2}  =\sqrt{ \frac{1}{2}\left( {A \pm \sqrt {A^2  - 4B} }
\right)},
\end{equation}
where $A = \alpha^2  + \beta^2  - 2\gamma^2$ and $B = \left( {\alpha
\beta - \gamma^2 } \right)^2 $. Thus the entropy of $\rho_{AB}$ is
\begin{equation}\label{24}
S(\rho _{AB} ) = g\left( {\lambda _1 } \right) + g\left( {\lambda _2
} \right),
\end{equation}
where
\begin{equation}\label{12}
g(x) = \frac{{x + 1}}{2}\log _2 \left( {\frac{{x + 1}}{2}} \right) -
\frac{{x - 1}}{2}\log _2 \left( {\frac{{x - 1}}{2}} \right).
\end{equation}
Let
\begin{equation}\label{40}
\begin{array}{l}
 C = [A\chi _d^2  + 2\alpha \left( {\alpha \beta  - \gamma ^2 } \right)\chi _d  + 2\gamma ^2  + B + 1]\left( {\beta  + \chi _d } \right)^{ - 2} , \\
 D = \left( {\alpha  + \sqrt B \chi _d } \right)^2 \left( {\beta  + \chi _d } \right)^{ - 2} , \\
 \end{array}
\end{equation}
then the entropy of $\rho^b_{AGH}$ is \cite{24}
\begin{equation}\label{32}
S\left( {\rho _{AGH}^b } \right) = g\left( {\lambda _3 } \right) +
g\left( {\lambda _4 } \right),
\end{equation}
where
\begin{equation}\label{41}
\lambda _{3,4}  = \sqrt {{1 \over 2}\left( {C \pm \sqrt {C^2  - 4D}
} \right)}.
\end{equation}
Since $S\left( {\rho _{AGH}^b } \right) $ is independent of Bob's
measurement $b$, we obtain
\begin{equation}\label{33}
\int {p(b)S\left( {\rho _{AGH}^b } \right)db}  = S\left( {\rho
_{AGH}^{b_0 } } \right),
\end{equation}
where $b_0$ is a constant.

So, in the discretely modulated CV-QKD, the lower bound to the
secret key rate is
\begin{equation}\label{15}
\tilde K=\beta\log _2 \left( {\frac{{V + \chi_t}}{{\chi_t + 1}}}
\right)-g\left( {\lambda _1 } \right) - g\left( {\lambda _2 }
\right) + g\left( {\lambda _3 } \right)+ g\left( {\lambda _4 }
\right).
\end{equation}
It is enough to derive the lower bound of the secret key rate
against collective attacks, because they are proven to be the most
powerful attacks in the asymptotic limit \cite{16,17}.

\section{IMPLEMENTATION OF DISCRETELY MODULATED CV-QKD}
\subsection{EXPERIMENTAL SETUP}
\begin{figure} \centering
\includegraphics[width=.5\textwidth]{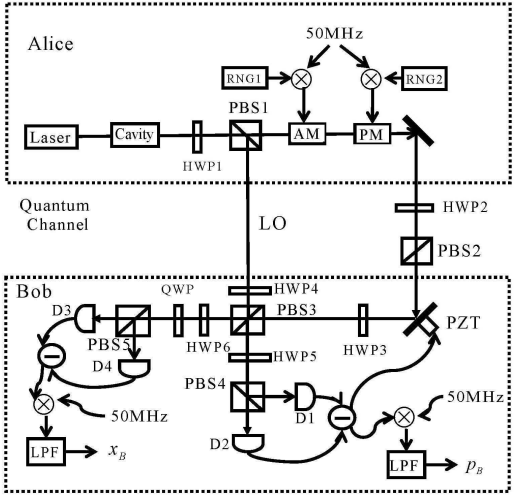}
\caption{The experimental schematic of discretely modulated CV-QKD.
Laser, NP Photonics (1550 nm); RNG1,-2, random number generators;
HWP1¨C6, half-wave plates; PBS1¨C5, polarizesr; PZT, piezoelectric
transducer; QWP, quarter-wave plate; D1¨C4, detectors; LPF, low-pass
filter.}\label{pic2}
\end{figure}

The schematic of our experimental setup is shown in Fig. \ref{pic2}.
The laser source is a 1550-nm continuous-wave fiber laser (NP
Photonics). The noise in the laser has been mainly suppressed by a
mode cleaner, which is a triangle resonant cavity with a finesse of
500.
Alice uses HWP1 and PBS1 to split a small part of the light as the
signal, and the rest as the LO. The signal's power is about 4$\mu$W,
and the LO's power is 40mW in our experiment. Then, Alice mixes two
random electronic signals with a 50-MHz carrier, and sends the
outputs of the mixers to amplitude and phase modulators to modulate
the signal. Alice generates binary pseudorandom numbers by using a
programmable function generator (Agilent33250A)in our experiment.
Since the generation rate of random electronic signals on Alice's
side is 10MHz, the width of each coherent state is 100ns. The LO and
the signal are sent to Bob through the quantum channel. We use HWP2
and PBS2 to replace the lossy channel (no excess noise).

When Bob receives the signal that Alice sends, he splits it into two
beams with HWP3 and PBS3. At one of the output ports of PBS3, he
makes a homodyne detection and uses the DC part of the result as a
feedback signal to control the PZT, so as to lock the relative phase
between the signal and the LO at ${\pi \mathord{\left/
 {\vphantom {\pi  2}} \right.
 \kern-\nulldelimiterspace} 2}$. So, he is actually measuring the $P$ quadrature at this port. At the other output port, he uses QWP to induce a $ {\pi  \mathord{\left/
 {\vphantom {\pi  2}} \right.
 \kern-\nulldelimiterspace} 2}$ phase shift between the signal and the LO, and he makes a homodyne detection to measure the $X$ quadrature.
We design a broadband balanced detector with a photodiode G8376-05
by Hamamatsu. The effective bandwidth is over 100 MHz, and the SNR
is near 9.2 dB for 20-mW coherent light as shown in Fig. \ref{pic3}.
The outputs of detectors are mixed with a 50-MHz carrier and
filtered by 25-MHz LPFs. The outputs of filters are sampled by a
data acquisition card NI PXIe-5122, and the sampling rate is 50 MHz.

\subsection{MODULATION AND NOISE SUPPRESSION}
In our experiment, the signal is not a weak coherent state but has a
large offset. When the signal and the LO's phase is locked, the
initial state can be written as $ \left| {x_0 } \right\rangle $,
while the modulated state can be written as $\left| {x_0 + x + ip}
\right\rangle $, where $x$ and $p$ are the signals added to the $X$
and $P$ quadratures of the light respectively. If $x,p \ll x_0 $,
amplitude $A$ and phase $\theta$ of the modulated light are
\begin{equation}\label{15}
\begin{array}{l}
 A = \sqrt {(x_0  + x)^2  + p^2 }  \approx x_0  + x,\\
\theta  = \arctan \left( {{p \mathord{\left/
 {\vphantom {p {\left( {x + x_0 } \right)}}} \right.
 \kern-\nulldelimiterspace} {\left( {x + x_0 } \right)}}} \right) \approx {p \mathord{\left/
 {\vphantom {p {x_0 }}} \right.
 \kern-\nulldelimiterspace} {x_0 }}.\\
\end{array}
\end{equation}
So, when Alice modulates the amplitude and the phase of the light,
she is just modulating the $X$ and $P$ quadratures of the light,
respectively. The half-wave voltage of the amplitude and the phase
modulators is 360V, while the electronic signals Alice adds to them
are less than 2 V, so the condition $x,p \ll x_0 $ is satisfied.

\begin{figure} \centering
\includegraphics[width=.5\textwidth]{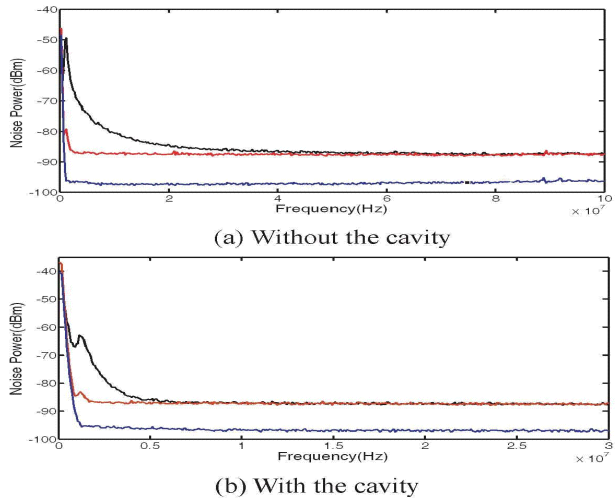}
\caption{(Color online) (a) The noise spectrum (0-100 MHz) of the
$X$ quadrature measurement of the signal without cavity (the highest
curve), shot noise (the middle curve), and electronics noise (the
lowest curve). (b) The noise spectrum (0-30 MHz) of the $X$
quadrature measurement of the signal with cavity (the highest
curve), shot noise (the middle curve), and electronics noise (the
lowest curve).}\label{pic3}
\end{figure}

The light generated by the fiber laser, which has much relative
intensity noise and phase noise in our experiment, can be treated as
a coherent state whose amplitude and phase are randomly modulated.
It is not a pure state but a mixed state and can be written as $\rho
= \int {P(n_A ,n_P )\left| {\alpha (n_A ,n_P )} \right\rangle
\left\langle {\alpha (n_A ,n_P )} \right|dn_A dn_P }$, where
${\left| {\alpha (n_A ,n_P )} \right\rangle }=\left| {\left( x_0 +
n_A \right)e^{in_P } } \right\rangle$. This can be regarded as the
fact that the source generates a pure state ${\left| {\alpha (n_A
,n_P )} \right\rangle }$ with the probability $ {P(n_A ,n_P )}$. So
for a particular state, we treat it as a coherent state $\left|
\alpha\left(n_A,n_P\right)\right\rangle$, but the parameters $n_A$
and $n_P$ are unknown. When Alice modulates its amplitude and phase
by $x$ and $p/x_0$, respectively, it becomes $\left| {\left( x_0 +
n_A+x \right)e^{i\left(n_P+p/x_0\right) } } \right\rangle $ and can
be written as $ \left| {\left( x_0 + n_A+x+ip \right)e^{in_P } }
\right\rangle$ when $x,p \ll x_0 $. The LO can be written as $
\left| {\alpha _{LO} e^{i\left( {\varphi + n_P } \right)} }
\right\rangle $, where $\alpha_{LO}$ is a real number, ${\alpha
_{LO}^2 }$ is the intensity of the LO, and $\varphi$ is the phase
shift added by Bob. The LO is quite strong, so its amplitude noise
can be ignored, and it has the same phase noise $n_P$ with the
signal when their optical path difference is far less than the
coherence length, since they come from one beam. In homodyne
detection, the difference of photoelectrons generated by two
photodiodes satisfies \cite{19}
\begin{equation}\label{17}
\Delta N_e
 \propto \alpha_{LO} \left( {\cos \varphi \hat x + \sin \varphi \hat p}
\right).
\end{equation}
From Eq. (\ref{17}), we can see that the phase noise in the laser
source has no effect on the results of the $P$ quadrature
measurement, since the phase noise $n_P$ does not appear in Eq.
(\ref{17}). This is consistent with our experimental results.
Whether the mode cleaner is added or not, the noise on the $P$
quadrature can reach the shot-noise limit. However, matters are
quite different with the $X$ quadrature. When we measure the noise
of the $X$ quadrature without the mode cleaner, there is a very
sizable classical noise with a sideband frequency up to 50 MHz, as
shown in Fig. \ref{pic3}(a). So, Alice uses a mode cleaner to purify
the noisy coherent state. After purification, the amplitude noise
above 10 MHz is remarkably suppressed and almost reaches the
shot-noise limit, as shown in Fig. \ref{pic3}(b).

Although the high-frequency noise above 10 MHz has been suppressed,
there is still much residual low-frequency noise under 10 MHz. In
order to avoid this noise, Alice uses two strings of random numbers
that she wants to send to mix with a 50-MHz carrier. Then, she
modulates the two mixed electronic signals to the amplitude and
phase modulators, respectively. Bob can filter the low frequency
noise easily and can pick up the interested frequency component by
using a mixer and an LPF. Finally, the noise in the source is
suppressed to the shot-noise limit and can be neglected. The
relative phase between the signal and the LO is locked, so the phase
noise of the interferometer is small enough to be ignored. In
Gaussian-modulation-based protocols, the modulation variance is
large, so it will induce notable excess noise \cite{9}. However, in
our experiment, the modulation variance is quite small (see the
following), and the noise caused by the imperfect modulation can be
ignored.

\subsection{DATA PROCESSING}
\begin{figure} \centering
\includegraphics[width=.5\textwidth]{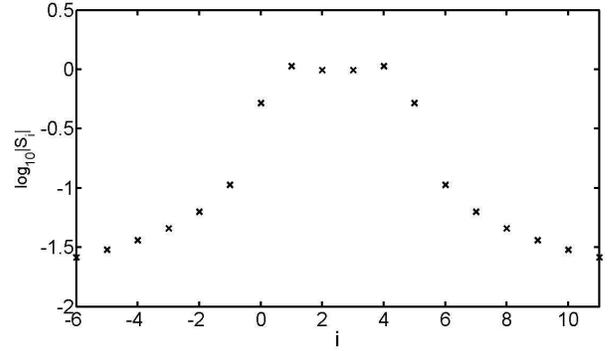}
\caption{$\log_{10} \left| {S_i } \right|$ as a function of
$i$.}\label{pic7}
\end{figure}

In order to get the measurement results of $X$ and $P$ quadratures
of the $n$th coherent state, Bob needs to get the $n$th difference
of photoelectrons $\Delta N_{en}$. Since the width of each coherent
state is $T = 100ns$, we obtain
\begin{equation}\label{11}
\Delta N_{en}  \propto \int_{(n-1)T}^{nT} {V(t)dt},
\end{equation}
where $V(t)$ is the output of the LPF. In practice, Bob does not
integrate $V(t)$, instead he samples it at $t_i=i\tau$ and gets
$V_i$ $(i = 0,1,2, \cdots )$, where $\tau  = 20ns $ is the sampling
interval, and $V_i$ is the sampling value. The bandwidth of $V(t)$
is less than 25MHz, so Bob can get all of the information of $V(t)$
with the sampling rate of 50MHz. Bob can rebuild $V(t)$ with his
samples \cite{20}
\begin{equation}\label{14}
V(t) = \sum\limits_{i =  - \infty }^\infty  {V_i {\rm{sinc}}} \left
(\frac{t}{\tau} - i \right ),
\end{equation}
so
\begin{equation}\label{16}
\begin{array}{l}
 \Delta N_{en}  \propto \int_{5(n - 1)\tau }^{5n\tau } {\sum\limits_{i =  - \infty }^\infty  {V_i {\rm{sinc}}\left(\frac{t}{\tau} - i\right)dt} }  \\
 {\kern 1pt} {\kern 1pt} {\kern 1pt} {\kern 1pt}{\kern 1pt} {\kern 1pt} {\kern 1pt} {\kern 1pt} {\kern 1pt} {\kern 1pt} {\kern 1pt} {\kern 1pt} {\kern 1pt} {\kern 1pt} {\kern 1pt} {\kern 1pt} {\kern 1pt} {\kern 1pt} {\kern 1pt} {\kern 1pt} {\kern 1pt}  = \sum\limits_{i =  - \infty }^\infty  {V_i \int_{(5n - i - 5)\tau }^{(5n - i)\tau } {{\rm{sinc}}\left(\frac{t}{\tau}\right)dt} } = \sum\limits_{i =  - \infty }^\infty  {V_i S_{i-5n+5} } , \\
 \end{array}
\end{equation}
where $ S_i  = \int_{- i\tau }^{(5- i)\tau } {{\rm{sinc}}(t/\tau)dt}
$ is symmetric with $i=2.5$. By considering that the absolute value
of $S_i$ attenuates quickly with the absolute value of $i-2.5$, as
shown in Fig. \ref{pic7},
in practice, the sum in Eq. (\ref{16}) is truncated from $i=0$ to
$i=5$. So, the measurement result of the $n$th coherent state (take
the $X$ quadrature for instance) is
\begin{equation}\label{19}
X_n \approx k\sum\limits_{i = 0}^5 {V_{5n + i-5} S_i } ,
\end{equation}
where the constant $k$ contains all the dimensional prefactors.

Thus, Bob approximately obtains the total difference of
photoelectrons when he measures the received signal  with this
method. If Eve attacks by measuring the signal during time windows
which are not sampled by Bob, she can not do it without being
discovered, because she will inevitably the total difference of
photoelectrons obtained by Bob. Actually, if Eve uses this attack,
she will increase the loss observed by Bob. So, the proposed
protocol is secure against this attack.

\subsection{EXPERIMENTAL RESULT AND DISCUSSION}
\begin{figure}
\centering
\includegraphics[width=.5\textwidth]{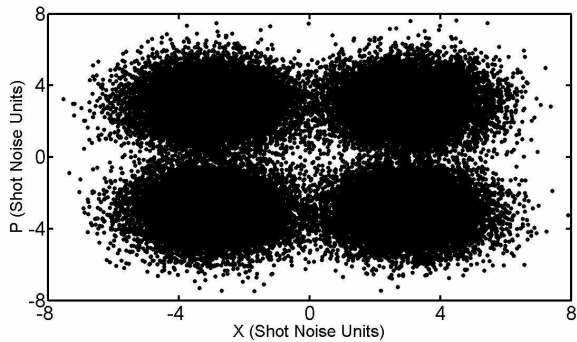}
\caption{The result of Bob's heterodyne measurement when the
modulation variance is 18.}\label{pic4}
\end{figure}

\begin{figure} \centering
\includegraphics[width=.5\textwidth]{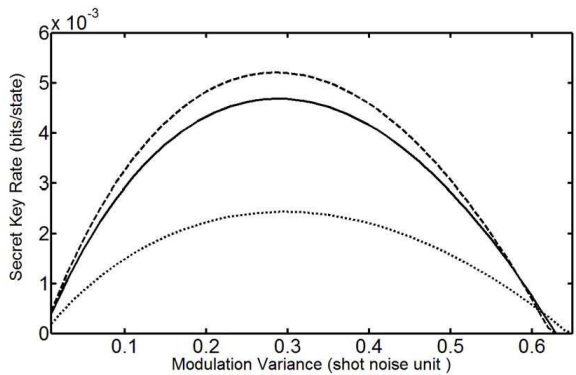}
\caption{The secret key rate as a function of the modulation
variance for an electronic noise of 0.12 (solid line), 0 (dashed
line), and 1.2 (dotted line). The channel is noiseless, and its
transmittance is 0.1; the efficiency of the detector is 0.8, and the
reconciliation efficiency is 0.8.}\label{pic5}
\end{figure}

The homodyne detectors are carefully calibrated, and their
efficiencies are both $\eta=0.8$. Then we need to determine the
excess noise in Bob's homodyne detector. We just simulate the
channel loss by using a beam splitter on a tabletop, so the
channel's excess noise $\varepsilon$ is 0. When the channel's
transmittance is $T=1$ and the modulation variance is 18, the result
Bob gets is shown in Fig. \ref{pic4}. There are 50000 points in this
figure. From the data, we can calculate the excess noise in the
detectors.
The total added noise is determined experimentally to be about
$\chi_t=1.8$. Since $T=1$, $\varepsilon=0$ and $\eta=0.8$, from Eq.
(\ref{18}) we obtain $\upsilon= 0.12$.  According to Eq. (\ref{2}),
we can find the maximal secret key rate by scanning the modulation
variance $V_a$, as shown in Fig. \ref{pic5}.
As a result, for a 90$\%$ lossy channel and the reconciliation
efficiency of 80$\%$, when the modulation variance is 0.29, we
achieve the maximal secret key rate of 46.8 kbits/s at the encoding
rate of 10 MHz.

We can also see that even if the electronic noise is suppressed at
0, the optimal secret key rate per coherent state is just $ 5.02
\times 10^{ - 3}$ bits, which is only a little bit higher than that
in our experiment. So, it is not very wise to enhance the secret key
rate by suppressing the electronic noise of the detectors.

In our experiment, the encoding rate is limited by the bandwidth of
the detectors, which is 100M. The secret key rate can be further
enhanced by increasing the encoding rate, which needs a broader
bandwidth detector and will lead to higher electronic noise. For
detectors with a certain gain-bandwidth product, if we broaden the
bandwidth $B$ to $\sqrt{10} B$, the gain $G$ becomes $10^{-0.5}G$,
which leads to a electronic noise of 1.2. In this case, the optimal
secret key rate is $2.43 \times 10^{ - 3} {{{\rm{bits}}}
\mathord{\left/
 {\vphantom {{{\rm{bits}}} {{\rm{state}}}}} \right.
  \kern-\nulldelimiterspace} {{\rm{state}}}}$, and the secret key
rate per second is 1.64 times as before. So, we can enhance the
secret key rate by decreasing the gain of the detectors so as to
broaden their bandwidth and to increase the encoding rate.

\section{CONCLUSION}
In this paper, we derive a lower bound to the secret key rate when
considering the noise in the source, and we present a discretely
modulated CV-QKD system by using strong coherent states and
heterodyne detection. We assume that the noise in the source is
induced by a neutral person Fred, and we present the equivalent
entanglement-based scheme. In this scheme, Eve can not purify the
state shared by Alice and Bob, so we can not calculate the secret
key rate. Instead, we derive a lower bound for the secret key rate.
In our experiment, the noise of the laser is suppressed at the
shot-noise limit by using a cavity and the method of frequency
shift. Since the modulation variance is quite small and the relative
phase between the signal and the LO is locked to perform heterodyne
detection, the excess noise induced by the imperfection of
modulation and the phase noise in the interferometer is negligible.
In order to increase the repetition rate, we broaden the bandwidth
of the detector at the expense of low SNR. When the channel loss is
90\%, we achieve a secret key rate of 46.8 kbits/s with the optimal
modulation variance and the encoding rate of 10 MHz. For detectors
with a certain gain-bandwidth product, the secret key rate can
further be improved by broadening the bandwidth of detectors and by
increasing the encoding rate. However, the secret key rate is only
an estimation and is not a real one. Many actions must be conducted,
such as error correction, privacy amplification, and reduction of
guided acoustic-wave Brillouin scattering in fiber-optic
implementations, which are what we will perform in a future work.

\section{ACKNOWLEDGMENTS}
The authors thank Hong Guo's group for useful discussions. The work
is supported by the key project of the Preparatory Research
Foundation of the National University of Defense Technology (Grant
No. JC08-02-01) and the National Natural Science Foundation of China
(Grant No. 10904174).

\end{document}